\documentstyle[aps,graphicx]{revtex}

\begin{document}

\title{ Experimental realization of
the Br\"{u}schweiler's algorithm in a homo-nuclear system
\thanks{Published in The
Journal of Chemical Physics 117, 3310 (2002).}}

\author{Li Xiao$^{a,b}$,
G. L. Long$^{a,b,c,d}$\thanks{corresponding author,
gllong@tsinghua.edu.cn},
Hai-Yang Yan$^{a,b}$,
Yang Sun$^{e,a,b}$}

\address{$^a$Department of Physics, Tsinghua University, Beijing 100084, China\\
$^b$ Key Laboratory For Quantum Information and Measurements,
Beijing 100084, China\\
$^c$ Center for Atomic and  Molecular Nanosciences, Tsinghua
University, Beijing 100084, China\\
$^d$Institute of Theoretical Physics, Chinese Academy of
Sciences, Beijing 100084, China\\
 $^e$Department of Physics and Astronomy, University of
Tennessee, Knoxville, TN 37996, U.S.A.} \maketitle

\begin{abstract}

Compared with classical search algorithms, $Grover$ quantum algorithm [$%
Phys. ${\it \ }$Rev.${\it \ }$Lett.${\it ,} {\bf 79}, 325(1997)] achieves
quadratic speedup and $Br\ddot{u}schweiler$ hybrid quantum algorithm [$Phys.$%
{\it \ }$Rev.$ $Lett.$, {\bf 85}, 4815(2000)] achieves an
exponential
speedup. In this paper, we report the experimental realization of the $Br%
\ddot{u}schweiler$ algorithm in a $3$-qubit NMR ensemble system.
The pulse sequences are used for the algorithms and the
measurement method used here is improved on that used by
Br\"{u}schweiler, namely, instead of quantitatively measuring the
spin projection of the ancilla bit, we utilize the shape of the
ancilla bit spectrum. By simply judging the downwardness or
upwardness of the corresponding peaks in an ancilla bit spectrum,
the bit value of the marked state can be read out, especially,
the geometric nature of this read-out can make the results more
robust against errors.
\end{abstract}


\section{Introduction}

Quantum algorithms are very important in quantum computing. One can find
this point in Deutsch and Josza's quantum algorithm which demonstrates the
incomparable advantage of quantum computing \cite{r3}. Two more famous
quantum algorithms which are closely related to practical applications of
quantum computation are: Shor's factoring algorithm \cite{r1} and Grover's
quantum search algorithm\cite{r2}. The factorization of a large number into
prime factors is a difficult mathematical problem because existing classical
algorithms require exponential times to complete the factorization in terms
of the input. However, Shor's quantum algorithm drastically decreases this
to polynomial times. Another similar example is searching marked items from
an unsorted database. Actually, many scientific and practical problems can
be abstracted to such search problem. Hence it is a very important subject.
Classically, it can only be done by exhaustive searching. Unlike Shor's
algorithm, Grover's quantum algorithm achieves only quadratic speedup over
classical algorithms, namely the number of searching is reduced from $O(N)$
to $O(\sqrt{N})$. However, it has been proven that Grover's algorithm is
optimal for quantum computing\cite{zalka99}. The strong restriction of the
optimality theorem can be broken off  if we go out of quantum computation
and then exponential speedup may be achieved. Using nonlinear quantum
mechanics, Abram and Lloyd\cite{abramlloyd} have constructed a quantum
algorithm that achieves exponential speedup. However the applicability of
nonlinear quantum mechanics is still under investigation, let alone the
realization of their algorithm at present.

Recently, by using multiple-quantum operator algebra, Br\"{u}schweiler put
forward a hybrid quantum search algorithm that combines DNA computing idea
with the quantum computing idea \cite{r4}. The new algorithm achieves an
exponential speedup in searching an item from an unsorted database. It
requires the same amount of resources as effective pure state quantum
computing. There are several known schemes for quantum computers, such as
cooled ions\cite{r5}, cavity QED\cite{r6}, nuclear magnetic resonance\cite
{r7} and so on. NMR technique is sophisticated and many quantum algorithms
have been realized by using NMR system\cite{r8,r9,r10,r11,r12,r13,rlong}.
Many studies show that the NMR system is particularly suitable for the
realization of such algorithms, in which ensembles of quantum nuclear spin
system are involved. Strictly speaking, Br\"{u}schweiler algorithm is not a
pure quantum algorithm, and thus the realization of Br\"{u}schweiler
algorithm in NMR enjoys the freedom from the debate\cite{rschack,rlaf} about
the quantum nature of the NMR computation using effective pure state.
Because this algorithm is exponentially fast, it tales much shorter time to
finish a search problem and this also makes the algorithm more robust again
decoherence.

In this paper, we report the experimental realization of
$Br\ddot{u}schweiler $ algorithm in a 3-qubit homo-nuclear
system. In the procedure, we have improved the measurement method
used by Br\"{u}schweiler in his paper\cite {r4}. Instead of
measuring the ancilla bit's spin polarization, we utilize the
shapes of ancilla bit's spectra, $i.e.$ by judging the
downwardness or upwardness of the corresponding peaks in the
spectrum, the bit value of the marked state can be read out.
Since the geometric property of the spectrum is easy to be
recognized, this makes the algorithm more tolerant to errors. Our
paper is organized as follows. After this introduction, we briefly
describe Br\"{u}schweiler's original algorithm in section
\ref{s2}, and then we introduce our modification part based on
$Br\ddot{u}schweiler$ algorithm in section \ref{s3}. In section
\ref{s4}, we present the details of the pulse sequences of the
algorithm and the results of our experiment. Finally, a summary
is given.

\section{Br\"{u}schweiler's algorithm}

\label{s2}NMR techniques lie far ahead of other suggested quantum
computing technologies. However, during recent years, the rapid
developing tendency becomes slow and slow. People have taken more
effort in preparing an effective pure state, but compared with a
pure state quantum computer, there is no essential speedup.
Br\"{u}schweiler's wonderful idea may shed a light on this area,
he takes advantage of the mixed state nature in the NMR system
and achieves an exponential speedup in searching an unsorted
database. For convenience in following discussion, here we repeat
the main idea of the Br\"{u}schweiler's algorithm in brief (in
detail, see Refs. \cite{r4,r16}).

As is well-known, the preparation of the effective pure state is one of the
most troublesome part in a NMR quantum computing experiment. On the other
hand, the effective pure state also sets a restriction on the number of
qubits\cite{r14,r15}. The effective pure state is represented by the density
operator
\begin{equation}
\rho =(1-\varepsilon )2^{-n}{\bf \hat 1}+\varepsilon \left|
00...0\right\rangle \left\langle 00...0\right| .  \label{e1}
\end{equation}

At room temperature, under the high temperature approximation we have
\begin{equation}
\varepsilon =\frac{nhv}{2^nkT}.  \label{e2}
\end{equation}

In Eq. (\ref{e1}), the second term's contribution to the outcome is scaled
by the factor $\varepsilon $, which decreases exponentially with $n$,
namely, the number of qubit, but the first term has no contribution at all%
\cite{contribution}.

In NMR ensemble system, the state can be represented by density operators
which are linear combinations of direct products of spin polarization
operators \cite{r7,r16}. In a strong external magnetic field, the
eigenstates of the Zeeman Hamiltonian
\begin{equation}
|\phi _{in}\rangle=\left| 001...01\right\rangle \>=\left| \alpha
\alpha \beta ...\alpha \beta \right\rangle ,
\end{equation}
are mapped on states in the spin Liouville space
\begin{equation}
\sigma _{in}=\left| \phi \right\rangle \left\langle \phi \right| ={ I}%
_1^\alpha { I}_2^\alpha { I}_3^\beta ...{ I}_{n-1}^\alpha { I}%
_n^\beta ,  \label{sigmain}
\end{equation}
where
\begin{equation}
{ I}_k^\alpha =\left| \alpha ^k\right\rangle \left\langle \alpha
^k\right| ={\frac 12}({\bf 1}_k+2{I}_{kz})=\left[
\begin{array}{cc}
1 & 0 \\
0 & 0
\end{array}
\right] ,
\end{equation}
\begin{equation}
{ I}_k^\beta =\left| \beta ^k\right\rangle \left\langle \beta ^k\right| ={%
\frac 12}({\bf 1}_k-2{ I}_{kz})=\left[
\begin{array}{cc}
0 & 0 \\
0 & 1
\end{array}
\right] ,
\end{equation}
represent respectively spin up and spin down state of the spin. Usually, the
oracle or query is a computable function $f$: $f(x)=0$ for all $x$ except
for $x=z$ which is the item that we want to find out for which $f(z)=1$.
Usually, the oracle can be expressed as a permutation operation which is a
unitary operation $U_f$, implemented using logic gates \cite{r4}. In Br\"{u}%
schweiler algorithm, an extra bit(also called the ancilla bit) is
used and its state is represented by ${ I}_0$. The output of the
oracle is stored on the ancilla bit ${ I}_0$ whose state is
prepared in the $\alpha $ state at the beginning. The output of
$f$ can be represented by an expectation value of ${ I}_{0z}$ for
a pure state
\begin{equation}
f=F({ I}_0^\alpha \sigma _{in})={\frac 12}-Tr(U_f{ I}_0^\alpha
\sigma _{in}U_f^{+}{ I}_{0z}).
\end{equation}

If $\sigma _{in}$ happens to satisfy the oracle, then $I_0^\alpha
$ is changed to $I_0^\beta $. This gives the value of the trace
equal to $-1/2$, and hence $f$ equals to 1. The input of $f$ can
be an mixed state of the form $\rho =\sum_{j=1}^N\;{ I}_0^\alpha
\;\sigma _j$ where $\sigma _j$ is one of the form in Eq.
(\ref{sigmain}):
\begin{equation}
f=\sum\limits_{j=1}^NF({ I}_0^\alpha \;\sigma _j)=F(\sum\limits_{j=1}^N%
{ I}_0^\alpha \;\sigma _j)+{\frac{{N-1}}2}.  \label{oracle}
\end{equation}

The oracle is applied simultaneously to all the components in the
NMR ensemble. The oracle operation is quantum mechanical.
Br\"{u}schweiler put forward two versions of search algorithm. We
adopt his second version. The essential of the Br\"{u}schweiler
algorithm is as follow: suppose that the unsorted database has
$N=2^n$ number of items. We need $n$ qubit system to represents
these $2^n$ items. The algorithm contains $n$ oracle queries each
followed by an measurement: \newline (1) Each time, ${
I}_0^\alpha { I}_k^\alpha $ ($k=1,2,...,n$) is prepared. In fact,
the input state ${ I}_0^\alpha ...{\bf 1}...I_k^\alpha ...{\bf
1}...$ is a highly mixed state\cite{r16}. In the following text,
the identity operator will be omitted. This Liouville operator
actually represents the $2^{n-1}$ number of items encoded in
mixed state:
\begin{eqnarray}
{ I}_0^\alpha { I}_k^\alpha  &=&{ I}_0^\alpha ({ I}_1^\alpha
+{I}_1^\beta )({ I}_2^\alpha
+{ I}_2^\beta )... ({ I}_n^\alpha +{ I}_n^\beta )  \nonumber \\
&=&\sum_{\gamma _1,\gamma _2,...,\gamma _{k-1},\gamma _k,\gamma
_{k+1},...,\gamma _n=\alpha ,\beta }{ I}_0^\alpha { I}_1^{\gamma
_1}{ I}_2^{\gamma _2}...{ I}_{k-1}^{\gamma _{k-1}}{ I}_k^\alpha {
I}_{k+1}^{\gamma
_{k+1}}...{I}_n^{\gamma _n}  \nonumber \\
&=&\sum_{i_1,i_2,...,i_{k-1},i_{k+1},...,i_n=0,1}\left|
i_1i_2...i_{k-1}0i_{k+1}...i_n\right\rangle \>\left\langle
i_1i_2...i_{k-1}0i_{k+1}...i_n\right| .  \label{state}
\end{eqnarray}

This mixed state contains half of the whole items in the database. The $k$%
-th bit is set to $\alpha $. The other half of the database with $k$-th bit
equals to $\beta $(or 1) is not included. (2) Applying the oracle function
to the system. As seen in eq. (\ref{oracle}), the operation is done
simultaneously to all the basis states. If $k$-th bit of the marked state is
$0$, then the marked state is contained in eq. (\ref{state}). One of the $%
2^{n}$ terms in equation (\ref{state}) satisfies the oracle and
the oracle changes the sign of the ancilla bit from $\alpha $ to
$\beta $. If one
measures the spin of ancilla spin after the function $f$, the value will be $%
f=(2^{n}-1)\times 1/2+1/2-(2^{n}-2)\times 1/2=1$. If the $k$-th bit of the
marked state is 1, then the state (\ref{state}) will not contain the marked
item. Upon the operation of the function $f$, there is no flip in the
ancilla bit. A measurement on the ancilla bit's spin $I_{0z}$ will yield $%
f=1/2\times (2^{n}-1)+1/2-(2^{n})\times 1/2=0$. However, without
obtaining the value of $f$, we can know the marked state by
measuring the ancilla bit's spin. If one measures the spin of
ancilla spin after the oracle, the value will be
$(2^{n-1}-1)\times 1/2-1/2=N/4-1$ for the $k$-th bit of the
marked state being 0. If the $k$-th bit of the marked state is 1,
then the state (\ref{state}) will not contain the marked item.
Upon the operation of the oracle, there is no flip in the ancilla
bit. A measurement on the ancilla bit's spin $I_{0z}$ will yield
$1/2\times (2^{n-1})=N/4$. Therefore by measuring the ancilla
bit's spin, one actually reads out the $k$-th bit
of the marked state. (3) By repeating the above procedure for $k$ from 1 to $%
n$, one can find out each bit value of the marked state.

In the following, we give a simple example with $N=4$ for illustrating the
algorithm, and the example is realized in an experiment. The example is used
for demonstration. The advantage of the algorithm will be seen if the number
of qubit becomes large. Suppose the unsorted database with four items $%
\{00,01,10,11\}$ is represented by Zeeman eigenstates of the two spins ${ I}_1$%
, ${ I}_2$. The item $z=10$ is the one which we want. That is to
say, $f={\bf 1} $ for $z=10$, which is expressed as ${ I}_1^\beta
{I}_2^\alpha $. For the other three items, $\{00({ I}_1^\alpha {
I}_2^\alpha )$, $01({ I}_1^\alpha { I}_2^\beta )$,
 $%
11({ I}_1^\beta { I}_2^\beta )\}$, $f=0$. Function $f$ can be
realized by a permutation illustrated in Fig.{1} (similar as
Figure 2 in Ref. \cite{r4}). The extra qubit ${ I}_0^\alpha $ is
included in the permutation.

First we prepare a mixed state ${ I}_0^\alpha { I}_1^\alpha $,
which is the sum of ${I}_0^\alpha { I}_1^\alpha { I}_2^\alpha +{
I}_0^\alpha { I}_1^\alpha { I}_2^\beta $. Then the permutation
described in Fig.{1} is operated on this mixed state. Since the
first bit of the marked state is 1, the permutation will have no
effect on the ancilla bit because it is obviously that the state
${I}_0^\alpha { I}_1^\alpha $ will not contain the marked state.
${ I}_0^\alpha { I}_1^\alpha {I}_2^\alpha $, ${ I}_0^\alpha {
I}_1^\alpha { I}_2^\beta $ each contributes $1/2$ to the spin of
the ancilla bit. Upon measurement of the ancilla bit on its spin,
the intensity will be $2\times 1/2=1$ unit. That tells us that
the first bit of the marked item is 1( in state ${ I}_1^\beta $).
Secondly, we prepare another state, ${ I}_0^\alpha { I}_2^\alpha
={ I}_0^\alpha {I}_1^\alpha { I}_2^\alpha +{ I}_0^\alpha {
I}_1^\beta { I}_2^\alpha $. We get output ${I}_0^\alpha {
I}_1^\alpha { I}_2^\alpha +{ I}_0^\beta { I}_1^\beta {
I}_2^\alpha $ after the action of
permutation $f$. Measuring the spin of ancilla bit, we get $0$, since $%
{ I}_0^\alpha { I}_1^\alpha { I}_2^\alpha $ and ${ I}_0^\beta {
I}_1^\beta { I}_2^\alpha $ contribute to the spin measurement
equally but with opposite signs. Then this tells us that the
second bit is $0 $( in state ${ I}_2^\alpha $). After these two
measurement, we have obtained the marked state. In the actual
experiment, we have modified the measuring part of the algorithm.
We read out the bit values by looking at the shape of the ancilla
bit. It is more clearly and concise.

\section{Modification to the original algorithm}

\label{s3}Need not measure the ${ I}_0^z$, we can distinguish the
state of the
ancilla bit by the shape of its spectrum. Because different initial state $%
{ I}_0^\alpha { I}_k^\alpha $ has the same form, except for the difference in the $%
k$ subscript, it is natural that the spectrum ${ I}_0$ will have
similar shapes for ${ I}_0^\gamma \ { I}_{k_1}^\delta $ and ${
I}_0^\gamma { I}_{k_2}^\delta $. We use the shape of the spectrum
of the state ${ I}_0^\alpha { I}_k^\alpha $ as a reference where
$k=1,2\cdots $. First, the phase of ${ I}_0^\alpha { I}_1^\alpha $
is determined as making peaks of the spectrum up. In this NMR system, the $%
{ I}_0$ bit has $J$ coupling to both ${ I}_1$ and ${ I}_2$ and
there are only two peaks in the ${ I}_0$ spectrum for the state
${ I}_0^\alpha { I}_1^\alpha $, the localities of two peaks are
determined by the order of the nuclei, which is the first,
second$\cdots $ The ${ I}_0$ spectrum of ${ I}_0^\alpha {
I}_1^\alpha $ before the operation of the permutation $U_f$ is
given in Fig. 2(a). After the permutation operation, we measure
the spectrum of ${ I}_0$ in new states again. If the shape of the
spectrum is the same as the one before the oracle, i.e., two
peaks are up still, then the permutation operation has not
changed the state ${ I}_0^\alpha { I}_k^\alpha $, and this means
that ${ I}_k$ is 1, that is to say, the $k$-th bit value of the
marked states $z$ is $1$. If the $k$-th bit of the marked state
is 0, the ancilla bit will flip after the operation of the
permutation $U_f$. We can see from density matrices before and
after the query operation $U_f$. Before the query is evaluated on
the mixed state ${ I}_0^\alpha { I}_1^\alpha $, the density
matrix(apart from a multiple of the identity matrix and a scaling
factor) is
\begin{equation}
\rho _{01in}=\left(
\begin{array}{cccccccc}
0.5 & 0 & 0 & 0 & 0.5 & 0 & 0 & 0 \\
0 & 0.5 & 0 & 0 & 0 & 0.5 & 0 & 0 \\
0 & 0 & 0 & 0 & 0 & 0 & 0 & 0 \\
0 & 0 & 0 & 0 & 0 & 0 & 0 & 0 \\
0.5 & 0 & 0 & 0 & 0.5 & 0 & 0 & 0 \\
0 & 0.5 & 0 & 0 & 0 & 0.5 & 0 & 0 \\
0 & 0 & 0 & 0 & 0 & 0 & 0 & 0 \\
0 & 0 & 0 & 0 & 0 & 0 & 0 & 0
\end{array}
\right) ,
\end{equation}

After the query, the matrix at the acquisition is
\begin{equation}
\rho _{01out}=\left(
\begin{array}{cccccccc}
0.5 & 0 & 0 & 0 & 0.5 & 0 & 0 & 0 \\
0 & 0.5 & 0 & 0 & 0 & 0.5 & 0 & 0 \\
0 & 0 & 0 & 0 & 0 & 0 & 0 & 0 \\
0 & 0 & 0 & 0 & 0 & 0 & 0 & 0 \\
0.5 & 0 & 0 & 0 & 0.5 & 0 & 0 & 0 \\
0 & 0.5 & 0 & 0 & 0 & 0.5 & 0 & 0 \\
0 & 0 & 0 & 0 & 0 & 0 & 0 & 0 \\
0 & 0 & 0 & 0 & 0 & 0 & 0 & 0
\end{array}
\right) .
\end{equation}
When we measure the spectrum of ancilla bit ${ I}_0$, the left
peak, corresponding matrix element $51$ and the right peak,
corresponding the matrix element $62$, do not change. This
indicates that the shape of the spectrum does not change. As for
the second step, before the query is evaluated on the mixed state
${ I}_0^\alpha { I}_2^\alpha $, the outcome matrix is
\begin{equation}
\rho _{01in}=\left(
\begin{array}{cccccccc}
0.5 & 0 & 0 & 0 & 0.5 & 0 & 0 & 0 \\
0 & 0 & 0 & 0 & 0 & 0 & 0 & 0 \\
0 & 0 & 0.5 & 0 & 0 & 0 & 0.5 & 0 \\
0 & 0 & 0 & 0 & 0 & 0 & 0 & 0 \\
0.5 & 0 & 0 & 0 & 0.5 & 0 & 0 & 0 \\
0 & 0 & 0 & 0 & 0 & 0 & 0 & 0 \\
0 & 0 & 0.5 & 0 & 0 & 0 & 0.5 & 0 \\
0 & 0 & 0 & 0 & 0 & 0 & 0 & 0
\end{array}
\right) ,
\end{equation}
and after the query, the matrix becomes
\begin{equation}
\rho _{02out}=\left(
\begin{array}{cccccccc}
0.5 & 0 & 0 & 0 & 0.5 & 0 & 0 & 0 \\
0 & 0 & 0 & 0 & 0 & 0 & 0 & 0 \\
0 & 0 & 0.5 & 0 & 0 & 0 & -0.5 & 0 \\
0 & 0 & 0 & 0 & 0 & 0 & 0 & 0 \\
0.5 & 0 & 0 & 0 & 0.5 & 0 & 0 & 0 \\
0 & 0 & 0 & 0 & 0 & 0 & 0 & 0 \\
0 & 0 & -0.5 & 0 & 0 & 0 & 0.5 & 0 \\
0 & 0 & 0 & 0 & 0 & 0 & 0 & 0
\end{array}
\right) .
\end{equation}

The left peak( $(51)$ matrix element) does not change, but the right peak,
((72) matrix element) changes sign. Thus the right peak of the spectrum will
be downward.

This method of ''reading out'' the bit of the marked state is
effective. Since it depends on the shape of the spectrum, a
topological quantity, it is insensitive to errors as compared to
the quantitative measurement of the spin of the ancilla bit.

\section{The realization of the algorithm in NMR experiment}

\label{s4}We implemented Br\"{u}hweiler algorithm in a 3 qubit
homonuclear NMR system. The physical system used in the experiment is $%
^{13}C $ labeled alanine
$^{13}C^1H_3-^{13}C^0H(NH_2^{+})-^{13}C^2OOH$. The solvent is
$D_2O$. The experiment is performed in a Bruker Avance DRX500
spectrometer. The parameters of the sample were determined by
experiment to be: $J_{02}=54.2$Hz, $J_{01}=35.1$Hz and
$J_{12}=1.7$Hz. In the experiment, $^1H$ is decoupled throughout
the whole process. $^{13}C^0$, $^{13}C^1$ and $^{13}C^2$ are used
as the 3 qubits, whose state are represented by ${ I}_0$, ${ I}_1$, ${ I}_2$ respectively. $%
^{13}C^0$ is used as the ancilla bit and the result of the oracle
is stored on it, $^{13}C^1$ and $^{13}C^2$ are the second and
third qubit respectively. We assume the marked item is $10$.

Firstly, the state ${ I}_0^\alpha { I}_1^\alpha $ is prepared. It
is achieved by a sequence of selective and non-selective pulses,
and $J$-coupling evolution. We begin our experiment from thermal
equilibrium state. This thermal state is expressed as,
\begin{equation}
\sigma (0_{-})=I_z^0+I_z^1+I_z^2.  \label{thermal}
\end{equation}

The input state $I_0^\alpha I_1^\alpha$ can be written as ${1\over
2}({1\over 2}{\bf 1}+I^0_z+I^1_z+2 I^0_zI^1_z) $. The identity
operator does not contribute signals in NMR, and a scale factor
is irrelevant, thus $I_0^\alpha I_1^\alpha$ is equivalent to
$I^0_z+I^1_z+2 I^0_zI^1_z$.  The pulse sequence\cite{r12,r18,r19}
\begin{equation}
\left( \frac \pi 2\right) _y^2\Rightarrow Grad\Rightarrow \left( \frac \pi 4%
\right) _x^{0,1}\Rightarrow \tau \Rightarrow \left( \frac \pi
6\right) _{-y}^{0,1}\Rightarrow Grad  \label{pulseinitial}
\end{equation}
applied to the thermal state produces this input state
\begin{equation}
\sigma(0_{+})={\sqrt{6} \over  4}(I^0_z+I^1_z+2 I^0_z I^1_z).
\label{stateop1}
\end{equation}
However in our experiment only the spectrum of $I_0$ is needed,
and only $J$ coupling between qubit 0 and 1 is retained, a
simplified pulse sequence is actually used in the present
experiment to prepare an equivalent input state:
\begin{equation}
\left( \frac \pi 2\right) _y^0\Rightarrow \tau'%
\Rightarrow \left( \frac \pi 2\right) _x^0 \Rightarrow \left( \frac %
\pi 4\right) _{-y}^0\Rightarrow Grad.  \label{pulsesecond}
\end{equation}
Here the subscripts denote the directions of the radio frequency pulse, and
the superscripts denote the nuclei on which the radio frequency are
operated. Two numbers at the superscript mean that the pulse are applied
simultaneously to two nuclei(In actual experiment, the pulses are applied in
sequence. Because the duration of the pulse is very short, they can be
regarded as simultaneous). $Grad$ refers to applying gradient field. $\tau
=1/(2J_{01})$ or $\tau'=1/(4J_{01})$ is the free evolution time during which nuclear $%
^{13}C^2$ is decoupled. The second pulse sequence is operated
more easily, because only selective to ${ I}_0$ is considered.
Pulse sequence (\ref{pulsesecond}) transforms the thermal state
(\ref{thermal}) into
\begin{eqnarray}
\sigma(0_{+})={1\over 2} (I^0_z+I^1_z+2I^0_zI^1_z)+{1\over
2}I^1_z+I^2_z. \label{stateop2}
\end{eqnarray}
States (\ref{stateop1}) and (\ref{stateop2}) are equivalent,
because ${1\over 2}I^1_z$ and $I^2_z$ does not contribute to the
$I^0$ spectrum, and the scaling factor does not matter.

 The oracle, represented as a permutation $f$ is
applied to this initial state: ${ I}_0^\alpha \ { I}_1^\alpha $.
Then result of the oracle operation is stored on the ancilla bit
${ I}_0$, that is, the state of the $^{13}C^0$ indicates the state
of the first bit of the marked item. Specifically the expression
of the unitary operation corresponding to the permutation $f$ is
\begin{equation}
U_f=\left(
\begin{array}{cccccccc}
1 & 0 & 0 & 0 & 0 & 0 & 0 & 0 \\
0 & 1 & 0 & 0 & 0 & 0 & 0 & 0 \\
0 & 0 & 0 & 0 & 0 & 0 & 1 & 0 \\
0 & 0 & 0 & 1 & 0 & 0 & 0 & 0 \\
0 & 0 & 0 & 0 & 1 & 0 & 0 & 0 \\
0 & 0 & 0 & 0 & 0 & 1 & 0 & 0 \\
0 & 0 & 1 & 0 & 0 & 0 & 0 & 0 \\
0 & 0 & 0 & 0 & 0 & 0 & 0 & 1
\end{array}
\right) .
\end{equation}

The permutation $U_f$ can be completed using a sequence logic
gates given in Fig. 3. The left one is CNOT gate and the right
one is Toffoli gate. The pulse sequence can be found out in
Ref.\cite{r17,r18}. The pulse sequence will be very complex if we
write according to the network although it is very rigorous.
Since we assume that there is only one marked state and only the
spectrum of $I_0$ is needed, the function of the $U_f$ can be
realized by the pulse sequence shown below
\begin{equation}
\left( \frac \pi 2\right) _y^0\Rightarrow \tau \Rightarrow \left( \frac \pi 2%
\right) _x^0,
\end{equation}
where $\tau =1/(2J_{01})$. After the operation of the oracle, we
measure the spectrum of the ancilla bit. This pulse sequence
achieves the same result as that for the gate shown in Fig.3:
\begin{eqnarray}
\begin{array}{ccc}
I^\alpha_0I^{\alpha}_1=I^\alpha_0I^\alpha_1I^\alpha_2+I^\alpha_0I^\alpha_1I^\beta_2
& \rightarrow &
I^\alpha_0I^\alpha_1I^\alpha_2+I^\alpha_0I^\alpha_1I^\beta_2\\
I^\alpha_0I^{\alpha}_2=I^\alpha_0I^\alpha_1I^\alpha_2+I^\alpha_0I^\beta_1I^\alpha_2
& \rightarrow &
I^\alpha_0I^\alpha_1I^\alpha_2+I^\beta_0I^\beta_1I^\alpha_2\end{array}
\end{eqnarray}

Secondly, the initial state ${ I}_0^\alpha { I}_2^\alpha $ is
prepared. There are two ways to prepare this initial state. One
method is to use a pulse sequence as in Eq. (\ref{pulseinitial})
or (\ref{pulsesecond}) by exchanging 1 with 2 in the
superscripts. Another method is to use the swap operator in
Ref.\cite{r16}
\begin{equation}
\left( \frac \pi 2\right) _y^{1,2}\Rightarrow \tau _1\Rightarrow \left(
\frac \pi 2\right) _x^{1,2}\Rightarrow \tau _1\Rightarrow \left( \frac \pi 2%
\right) _{-y}^{1,2},  \label{swap}
\end{equation}
onto the initial state ${ I}_0^\alpha { I}_1^\alpha $ and the
state ${ I}_0^\alpha { I}_2^{\alpha }$ will be obtained. In the
experiment, we adopt the second approach.  The swap operator is
important in generalizing the experiment into more qubit system
and we will discuss this later. Then we apply the permutation
$U_f$ again, and the result of the oracle is stored in the
ancilla bit $^{13}C^0$.

The spectra for ${ I}_0 $ after the oracle query $U_f $ operated
on ${ I}_0^\alpha { I}_1^\alpha $ and ${ I}_0^\alpha {
I}_2^\alpha $ are given in Fig. 2 (b) and Fig. 2 (c)
respectively. We can see clearly that the one has the same shape
as the reference spectrum and the other one has flipped the right
peak. This tells us that the first bit and the second bit of the
marked state are $1$ and $0$ respectively. Thus the marked state
is $10$. We also notice that there are small differences between
the spectra before and after the permutation operations for ${
I}_0^\alpha{ I}_1^\alpha$. These are expected due to imperfections
caused by the inhomogeneous field, the errors in the selective
pulse and in the evolution of chemical shift.

\section{Summary}

\label{s5}In summary, we have successfully demonstrated the
Br\"{u}schweiler algorithm in a 3-qubit homo-nuclear NMR system.
Pulse sequences are given. A new method for ``reading out'' the
bit value of the marked state is proposed and realized. The
number of iteration required for this algorithm is very small.
This is particularly propitious to resist decoherence, especially
for NMR system at room temperature. Another advantage of this
algorithm is its robustness against errors, $i.e.$ the shape of
the spectrum in reading the bit of the marked state has a special
feature and one can easily distinguish it from others. Another
advantage is its high probability in finding the marked state, it
is 100\%!.

 It should be point out that there are still several
issues to be addressed in generalizing the searching machine to
more qubit system. First, one must
find a suitable molecule to act as the quantum computer. According to Br\"{u}%
schweiler's original algorithm, the ancilla qubit ${ I}_0^\alpha
$ must interact with every other qubit. However, in a molecule,
the interaction between remote nuclear spins is very weak. This
may be overcome by the swap operation as given in Ref.
\cite{r16}. Using swap operation, we can prepare any initial
state ${ I}_0^\alpha { I}_k^\alpha $ without the direct
interaction between spin ${ I}_0^\alpha $ and spin ${ I}_k^\alpha
$. And, the qubit also can be read out easily from the shape of
the spectrum. All these are under consideration in our future
work.

The authors thank Prof. X. Z. Zeng, Prof. M.L. Liu and Dr. J. Luo
for help in preparing the NMR sample and the use of selective
pulses. Helpful discussions with Dr. P. X. Wang are gratefully
acknowledged. This work is supported in part by China National
Science Foundation,  the National Fundamental Research Program,
Contract No. 001CB309308 ,
the Hang-Tian Science Foundation. \newpage {\Large %
References\bigskip \bigskip }

\begin{figure}
\begin{center}
\includegraphics[height=2in]{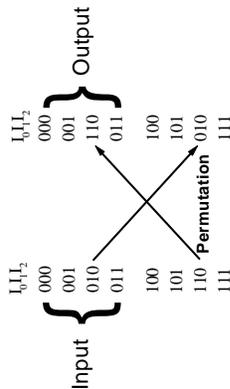}
\caption{Representation of the oracle $U_f$ spanning on spins
$I_1$, $I_2$ whose function corresponds to one permutation and
results of the query stored on $I_0$.}
\end{center}
\end{figure}

\begin{figure}
\begin{center}
\includegraphics[width=4in]{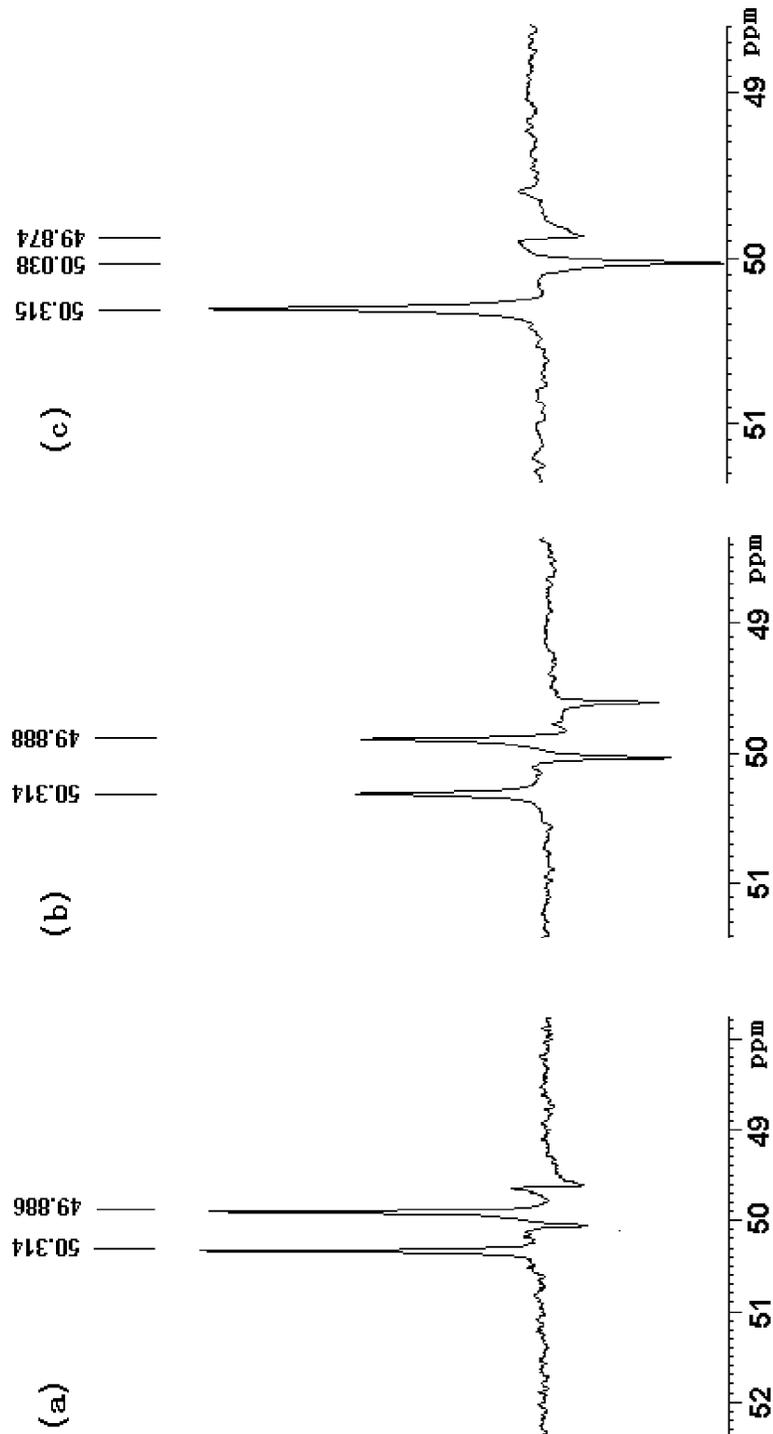}
\caption{Experimental realization of the Br\"{u}schweiler's
algorithm using
the three spins $1/2$ of $^{13}C$ , $^{13}C$-labled alanine dissolved in $%
D_2O$. (a) The spectrum of the ancilla qubit in state before the oracle $U_f$%
, two peaks are upword. The shape of the spectrum is used as
criterion. (b)
The spectrum of the ancilla qubit after the oracle $U_f$. The first bit is $%
1 $. (c) The spectrum of the ancilla qubit in the state after the oracle $%
U_f $. The second bit is $0$. The marked state is $10$.}
\end{center}
\end{figure}

\begin{figure}
\begin{center}
\includegraphics[height=2in,angle=-90]{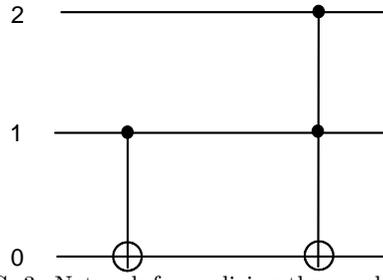}
\caption{Network for realizing the oracle $U_f.$}
\end{center}
\end{figure}

\end{document}